\documentclass[10pt, conference]{IEEEtran}

\usepackage[T1]{fontenc}
\usepackage[english]{babel}
\usepackage{graphicx}
\usepackage{float}
\usepackage{booktabs}
\usepackage{multicol}
\usepackage{multirow}
\usepackage[table,xcdraw]{xcolor}
\usepackage{amsmath, amsfonts, mathtools, amsthm, amssymb}
\usepackage{mathrsfs}
\usepackage[inline,shortlabels]{enumitem}
\usepackage[breaklinks = true, hidelinks]{hyperref}
\usepackage{lipsum}
\usepackage[utf8]{inputenc}
\usepackage{color,soul}
\usepackage[per-mode=symbol,range-phrase=~--~]{siunitx}
\usepackage{makecell}
\usepackage{tikz}
\usepackage{caption}
\usepackage{subcaption}

\newcommand{\deep}{\mbox{\texttt{DEEP}}}
\newcommand{\dcloudID}{\mathrm{dcloud2.itec.aau.at/aau}} %
\newcommand{\DockerHubID}{\mathrm{sina88}} %
\newcommand{\CORE}{\mathtt{CORE}}
\newcommand{\CPU}{\mathtt{CPU}}
\newcommand{\MEM}{\mathtt{MEM}}
\newcommand{\STOR}{\mathtt{STOR}}
\newcommand{\BW}{\mathtt{BW}}
\newcommand{\SIZE}{\mathtt{Size}}
\newcommand{\Data}{\mathit{df}}
\newcommand{\sched}{\mathtt{sched}}
\newcommand{\regist}{\mathtt{regist}}
\newcommand{\ct}{\mathtt{CT}}
\newcommand{\ec}{\mathtt{EC}}
\newcommand{\REQ}{\mathtt{req}}

\begin{document}

\title{DEEP: Edge-based Dataflow Processing with Hybrid Docker Hub and Regional Registries}

\author{Narges Mehran\IEEEauthorrefmark{1}
, Zahra Najafabadi Samani\IEEEauthorrefmark{3}, Reza Farahani\IEEEauthorrefmark{4}, Josef Hammer\IEEEauthorrefmark{5}, Dragi Kimovski\IEEEauthorrefmark{4}\\
\IEEEauthorblockA{\IEEEauthorrefmark{1} Intelligent Connectivity, Salzburg Research Forschungsgesellschaft mbH, Austria}
\IEEEauthorblockA{\IEEEauthorrefmark{3} Institute of Computer Science, University of Innsbruck, Austria}
\IEEEauthorblockA{\IEEEauthorrefmark{4} Institute of Information Technology, University of Klagenfurt, Austria}
\IEEEauthorblockA{\IEEEauthorrefmark{5} Institute for Network and Security, Eastern Switzerland University of Applied Sciences, Switzerland}
}

\maketitle
\begin{abstract}
Reducing energy consumption is essential to lessen greenhouse gas emissions, conserve natural resources, and help mitigate the impacts of climate change. In this direction, edge computing, a complementary technology to cloud computing, extends computational capabilities closer to the data producers, enabling energy-efficient and latency-sensitive service delivery for end users. 
To properly manage data and microservice storage, expanding the Docker Hub registry to the edge using an AWS S3-compatible \emph{MinIO}-based object storage service can reduce completion time and energy consumption. To address this, we introduce \emph{Docker rEgistry-based Edge dataflow Processing} (\emph{\deep{}}) to optimize the energy consumption of microservice-based application deployments by focusing on deployments from Docker Hub and MinIO-based regional registries and their processing on edge devices. After applying \emph{nash equilibrium} and benchmarking the execution of two compute-intensive machine learning (ML) applications of video and text processing, we compare energy consumption across three deployment scenarios: exclusively from Docker Hub, exclusively from the regional registry, and a hybrid method utilizing both. 
Experimental results show that deploying \SI{83}{\percent} of text processing microservices from the regional registry improves the energy consumption by \SI{.34}{\percent} ($\approx$\SI{18}{\joule}) compared to microservice deployments exclusively from Docker Hub.
\end{abstract}

\textcolor{red}{\scriptsize 2025 IEEE. Personal use of this material is permitted. Permission from IEEE must be obtained for reprinting/republishing/redistributing/reusing this work in other works.}
\section{Introduction}
Edge computing addresses the need to reduce energy consumption by extending computational capabilities closer to where the data is produced. This reduces the energy required for data transfers to centralized cloud servers, enabling more energy-efficient and latency-sensitive services for end users. Moreover, edge computing allows non-time-critical tasks to be offloaded to the cloud, balancing local processing and centralized resource optimization. This combination supports sustainable and efficient service delivery while contributing to global efforts to reduce energy use and environmental impact.

To effectively support edge computing, Docker Hub\footnote{\url{https://hub.docker.com/}\label{footnot:dock-hub}} serves as a widely used container registry for storing, managing, and sharing Docker images. It leverages a network of cloud data centers and content delivery networks (CDNs) to guarantee low latency and scalability for developers. While the locations of Docker Hub's servers remain undisclosed, its CDN-based distribution model enables Docker images to be served geographically closer to end users, reducing download times.
This 
model aligns with other large-scale cloud services, combining centralized servers for data management with distributed point of presence to enhance delivery performance. Moreover, recently, \emph{regional computing}~\cite{badshah2022use} was presented as a strategy for data processing and storage during \emph{on-peak network hours}, with the ability to migrate 
processed data to cloud servers during \emph{off-peak hours}, 
reducing network congestion. 

Traditional methods~\cite{peinl2016docker,zerouali2023helm} often rely on cloud-native Docker registries, which typically result in higher latency and energy consumption while overlooking the potential benefits of regional registries. Although some efforts have designed MinIO-based regional Docker registries~\cite{makris2024edge,hammer2023distributed} and caching mechanisms~\cite{khan2021cache,hua2020fog} to minimize download times, they rarely focus on incorporating energy-efficient microservice scheduling strategies. 
To solve these challenges, we introduce \deep{}, which investigates energy efficiency by addressing both the microservice image deployment from Docker registries and their dataflow processing on edge devices. \deep{} explores the potential of a regional Docker registry in boosting environmental sustainability and energy efficiency. 
More specifically, \deep{} applies a \emph{nash equilibrium} model to improve the energy consumption of compute-intensive ML applications, modeled as a directed acyclic graph (DAG) deployed from Docker registries. 
Therefore, the primary contributions of \deep{} are:
\begin{itemize}[leftmargin=*,align=left]
    \item Design and develop a regional registry for image retrieval;
    \item Formulate a nash-equilibrium game-based model for microservice deployment on edge devices; 
    \item Benchmark and compare the energy consumption of the microservice deployments from two Docker registries and executions on a physical edge-based testbed.
\end{itemize}
This paper has six sections. 
We review the related work in Section~\ref{sec:related}. 
Section~\ref{sec:model:arch} describes the models of application, devices, execution, the investigated problem, and the \deep{} architecture. Sections~\ref{sec:experimntdesign}--\ref{sec:experimntresult} present the experimental design and results before concluding the paper in  Section~\ref{sec:conclusion}.

\section{Literature Review}\label{sec:related}

\subsubsection*{Cloud-native Docker registry} 
Zerouali et al.~\cite{zerouali2023helm} investigated which source registries host the most frequently used images, such as Busybox, MySQL, Nginx, Postgres, and Alpine, with the expectation that most stem from Docker Hub, with over \num{10} million images. Their findings show that Docker Hub is the most popular registry, GitHub is the second most popular, and Quay~\cite{quay/quay} is the third most popular. 

\subsubsection*{Edge-driven regional Docker registry}
Makris et al.~\cite{makris2024edge} introduced a data placement method that minimizes completion time by leveraging an edge-driven Docker registry tailored for extended reality applications. 
With the development of augmented and virtual reality technologies, this method proposes an edge-located Docker registry that caches Docker images closer to the application's users, reducing reliance on cloud-native registries that may impose extra delays.

\subsubsection*{Edge-driven cache-based scheduling}
Khan et al.~\cite{khan2021cache} introduced a task allocation strategy in fog computing, which bridges the gap between cloud and edge devices by leveraging caching mechanisms to reduce latency. This scheme stores frequently accessed data closer to the edge, minimizing repeated data transfers 
and improving response times.  
Hua et al.~\cite{hua2020fog} proposed a data caching scheme that allocates fog resources closer to users by integrating proactive and reactive scheduling methods, enhancing resource efficiency and reducing latency. 

\subsubsection*{Energy-aware ML application deployment}
Mokhtari et al.~\cite{mokhtari2022felare} proposed a heuristics-based deployment method for ML applications on heterogeneous edge devices. This method minimizes task completion time while adhering to energy constraints for edge deployment, particularly latency-sensitive Internet of Things-based applications, e.g., SmartSight.

\subsubsection*{Research gap} Existing methods have focused on either solely cloud-native or regional registry~\cite{zerouali2023helm}-\cite{makris2024edge}, caching mechanisms~\cite{khan2021cache}-\cite{hua2020fog}, or  energy efficiency in machine learning applications deployment~\cite{mokhtari2022felare}. In contrast, this work explores energy efficiency in ML application deployment by leveraging both cloud-native and edge-based regional hybrid Docker registries.

\section{Model and Architecture}\label{sec:model:arch}
\subsection{Application model}\label{sec:appmodel}
\subsubsection*{Dataflow processing application} is a DAG $\mathcal{A} = \left(\mathcal{M}, \mathcal{E}\right)$ consisting of \emph{microservices} interdependent through \emph{dataflows}:
\begin{enumerate}[align=left,leftmargin=*]
\item \emph{Microservices} $\mathcal{M} =\left\{\left(m_i,\SIZE_{m_i}\right) | 0 \leq i < \mathcal{N}_{\mathcal{M}}\right\}$ where $\mathcal{N}_{\mathcal{M}}$ denotes the number of microservices in application $\mathcal{A}$ and $\SIZE_{m_i}$ is the \si{\giga\byte} size of each containerized $m_i$;
\item \emph{Dataflows} $\mathcal{E} = \{\left(m_{u}, m_{i}, \Data_{ui}\right)$ | $ \left(m_{u}, m_{i}\right) \in \mathcal{M} \times \mathcal{M}\}$ with $\SIZE_{ui}$ (in \si{\mega\byte}), transferring from an \emph{upstage} microservice $m_{u} \in \mathcal{A}$ to a \emph{downstage} one $m_{i} \in \mathcal{A}$.
\end{enumerate}
\paragraph*{Requirements} $\REQ\left(m_{i}\right)$ for processing dataflows $\Data_{ui}$ by a microservice $m_{i}$ is a triple representing the minimum number of cores, memory, and storage sizes (in \unit{\giga\byte}):
\begin{equation*}
    \REQ\left(m_{i}\right) = \langle\CORE\left(m_{i}\right),\CPU\left(m_{i}\right),\MEM\left(m_{i}\right),\STOR\left(m_{i}\right)\rangle,
\end{equation*}
where $\CPU\left(m_{i}\right)$ represents the minimum processing load, measured in millions of instructions (\unit{MI}), required to process dataflows $\Data_{ui}$ using $m_{i}$.
\subsection{Device and network model}\label{sec:devmodel}
We model the 
set of heterogeneous capacity-constrained devices interconnected via network channels:

\begin{enumerate}[align=left,leftmargin=*]
\item {Devices} $\mathcal{D} = \left\{d_j | 0 \leq j < \mathcal{N}_{\mathcal{D}}\right\}$ represent a set of $\mathcal{N}_{\mathcal{D}}$ physical edge  devices, each characterized by $\CORE_j$ \emph{cores}, $\MEM_j$ \emph{memory} and $\STOR_j$ \emph{storage} (in \si{\giga\byte}) and $\CPU_j$ \emph{processing speed} (in million of instructions per second (\si{MI\per\second})): \mbox{$d_j = \left(\CORE_j,\CPU_j,\MEM_j,\STOR_j\right)$}.
\item {Network channels} \mbox{$\mathcal{H}=\{h_{kj}\ |\ 0 \leq k,j < \mathcal{N}_{\mathcal{D}}\}$} interconnect the devices, where each channel $h_{kj}=\BW_{kj}$ models the network bandwidth $\BW_{kj}$ for transferring a data unit between the devices $d_k$ and $d_j$. For simplicity, this work neglects the network round-trip time (RTT), focusing exclusively on bandwidth for data transfer.
\end {enumerate}
\subsection{Registry model} 
We model the Docker image registries \mbox{$\mathcal{R}=\left\{r_g | 0 \leq g < \mathcal{N}_{\mathcal{R}}\right\}$} to store the containerized microservices $m_i\in\mathcal{M}$ where $\mathcal{N}_{\mathcal{R}}$ shows the number of registries. $\BW_{gj}$ represents the network bandwidth for transferring containerized microservices between the registry server $r_g$ and device $d_j$, hosting microservice $m_i$.
\subsection{Dataflow processing model} \label{sec:execmodel}
Dataflow processing consists of the following key metrics:
\begin{enumerate}[align=left,leftmargin=*]
\item{Completion time}\label{model:flowproctime}
$\ct\left(m_i,r_g,d_j\right)$ for a dataflow $\Data_{ui}$ processed by a microservice $m_i$, retrieved from a registry $r_g$, and scheduled on a device $d_j$, is the sum of
\emph{deployment}, \emph{dataflow transmission} and \emph{dataflow processing} times:
\begin{equation*}
\ct\left(m_i,r_g,d_j\right)=\frac{\SIZE_{m_i}}{\BW_{gj}}+ \frac{\SIZE_{ui}}{\BW_{kj}}+\frac{\CPU\left(m_i\right)}{\CPU_j},
\end{equation*}
where deployment time ($T_d$) is determined by downloading a containerized microservice $m_i$ of size $\SIZE_{m_i}$ not already existing on a device $d_j$ from a Docker registry (remote or regional) and deploying it on $d_j$ while $\BW_{gj}$ defines the bandwidth from Docker registry $r_g$ to a device $d_j$. The dataflow transmission time ($T_c$) is defined based on the transfer between microservices $m_u$ and $m_i$ executing on corresponding devices $d_k$ and $d_j$. Finally, the \mbox{dataflow processing time ($T_p$)} is computed as the ratio of the processing requirements to the device's capacity $\CPU_j$. Note that we utilized the application's DAG to model its completion time~\cite{samani2023incremental}.
\item{Energy consumption} $\ec\left(m_{i},r_g,d_j\right)$ is the sum of the \textit{active} and \textit{static} energy consumed $\ec\left(m_{i},d_j\right)$ during each microservice (non-concurrently) execution:
\begin{equation*}
\ec\left(m_{i},r_g,d_j\right) = 
E_a\left(m_{i},r_g,d_j\right) + E_s\left(d_j\right),
\end{equation*}
where $E_a\left(m_{i},r_g,d_j\right)$ represents \emph{active energy} consumed during microservice $m_i$ deployment from $r_g$ on device $d_j$ to process a dataflow $\Data_{ui}$, which is directly related to $\ct\left(m_i,r_g,d_j\right)$. Moreover, $E_s\left(d_j\right)$ denotes \emph{static energy} consumed for maintaining the device $d_j$ and running its background tasks. The total energy $\ec_\text{total}\left(\mathcal{A},\mathcal{R},\mathcal{D}\right)$ consumed by an application $\mathcal{A}$ is the sum of energy consumed by all its microservices during dataflow processing~\cite{mehran2019mapo}:
\begin{equation*}
\ec_{\text{total}}\left(\mathcal{A},\mathcal{R},\mathcal{D}\right) = \sum\limits_{\substack{m_i \in \mathcal{A}\ \land \\ d_j=\sched\left(m_i\right)\ \land\\
r_g=\regist\left(m_i\right)}} \ec\left(m_i, r_g, d_j\right).
\end{equation*}
\end{enumerate}
\subsection{Problem definition}\label{sec:prob}
We model the dataflow processing problem in the context of the Docker registry-based microservice image storage services using the \emph{prisoner dilemma} model within the nash equilibrium~\cite{gkasior2021distributed} to optimize energy consumption through cooperation between microservices and devices. The game investigates the solutions for retrieving the microservices from a Docker registry with minimal energy consumption during deployment from registries \mbox{$r_g=\regist\left(m_i\right)$} and executing them on the least power-hungry edge devices \mbox{$d_j=\sched\left(m_i\right)$}: 
\begin{align}
    \min\limits_{\substack{\sched(\mathcal{M}) = \mathcal{D}\ \land \ \regist(\mathcal{M}) = \mathcal{R}}}\ec_{\text{total}}\left(\mathcal{A},\mathcal{R},\mathcal{D}\right).\notag
    \label{eq:prob}
\end{align}
\subsection{Architecture}\label{sec:arch}
\begin{figure}[!t]
    \centering
    \includegraphics[width=0.7\columnwidth]{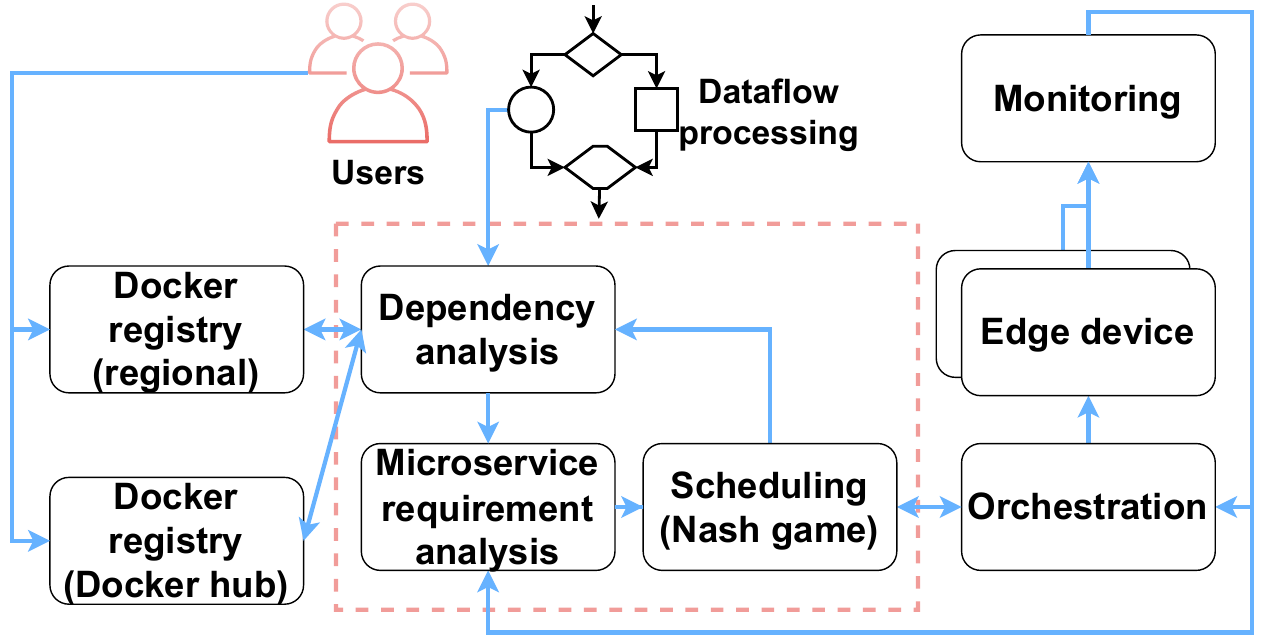}
    \caption{\deep{} 
    architecture.}
    \label{fig:arch}
\end{figure}
Figure~\ref{fig:arch} shows \deep{}'s architecture, consisting of three main components: dataflow dependency analysis, microservice requirement analysis~\cite{farahani2024heftless}, and nash-game-based scheduling on the computing infrastructures. In this architecture, the user instantiates the microservice deployment. Respectively, the dependency analysis and the scheduling are loosely coupled with Docker registries and an orchestrator, such as the open-source Kubernetes. A monitoring system logs the service executions on the computing devices.
\section{Experimental Design}\label{sec:experimntdesign} 
We implemented \deep{} in \mbox{\texttt{Python 3.9.13}} using the \texttt{Nashpy} library\footnote{\url{https://github.com/drvinceknight/Nashpy}} for solving 
dataflow processing problem.
\subsection{Computing devices} We created a testbed of two edge devices with \texttt{Intel} and \texttt{ARM} architectures:
\begin{enumerate*}
    \item \texttt{medium}-size device: \num{8}-core Intel$^\circledR$ Core$^{(TM)}$ \mbox{i7-7700} processor, \qty{16}{\giga\byte} of memory, and \qty{64}{\giga\byte} of storage running Ubuntu \num{20.04};
    \item \texttt{small}-size device: \num{4}-core ARM-based Raspberry Pi 4, with \SI{8}{\giga\byte} of memory and \SI{32}{\giga\byte} of storage running \texttt{Debian GNU/Linux 12 bookworm}.
\end{enumerate*}

\subsection{Application case studies}
\begin{figure}[!t]
    \centering
    \subfloat[Video processing]{\includegraphics[width=.45\columnwidth]{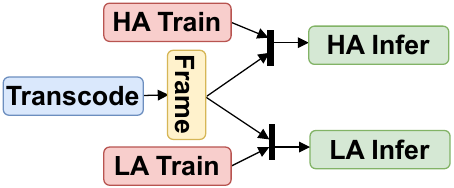}\label{fig:app1}}
    \hfill
    \subfloat[Text processing]{\includegraphics[width=.48\columnwidth]{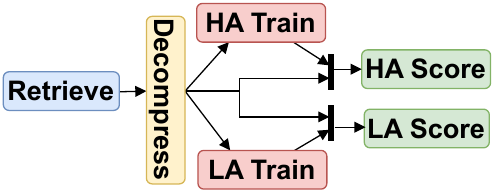}\label{fig:app2}}
    \caption{Case study applications.}
    \label{fig:apps}
\end{figure}
Figure~\ref{fig:apps} shows two microservice-based applications~\cite{farahani2024heftless}
designed and benchmarked on the physical testbed~\cite{samani2023incremental}:
\begin{enumerate*}
\item \emph{video processing} consisting of six microservices, including video transcoding received from a camera source in high quality, framing into still images, high-accuracy (HA) and low-accuracy (LA) training and inferring of the road sign object in the frame;
\item \emph{text processing} comprises six microservices consisting of retrieving the Amazon annotated reviews from a repository residing in an AWS S3 bucket, decompressing the dataset into training and testing parts, HA/LA training on the training dataset, and HA/LA scoring of the testing dataset.
\end{enumerate*}

Each application comprises two synchronization barriers defining the dependencies of a downstage microservice to its upstage ones for execution. 
\begin{table}[t]
\centering
\caption{Docker images of microservices.}
\label{tab:docker-img}
\resizebox{\columnwidth}{!}{
\begin{tabular}{|c||c|c|}
\cline{2-3}
\multicolumn{1}{c|}{}  & \emph{Docker Hub} & \emph{AAU Regional Registry}\\
\hline
\multirow{6}{.4cm}{\centering {\rotatebox[origin=c]{90}{\makecell{{{\textit{Video}}}\\{{\textit{processing}}}}}}}&$\DockerHubID$/vp-transcode &$\dcloudID$/vp-transcode\\
\cline{2-3}
&$\DockerHubID$/vp-frame &$\dcloudID$/vp-frame\\
\cline{2-3}
&$\DockerHubID$/vp-ha-train &$\dcloudID$/vp-ha-train\\
\cline{2-3}
&$\DockerHubID$/vp-ha-infer &$\dcloudID$/vp-ha-infer\\
\cline{2-3}
&$\DockerHubID$/vp-la-train &$\dcloudID$/vp-la-train\\
\cline{2-3}
&$\DockerHubID$/vp-la-infer &$\dcloudID$/vp-la-infer\\
\cline{2-3}
&$\DockerHubID$/vp-ha-infer &$\dcloudID$/vp-ha-infer\\
\hline
\multirow{6}{.4cm}{\centering {\rotatebox[origin=c]{90}{\makecell{{{\textit{Text}}} \\{{\textit{processing}}}}}}}&$\DockerHubID$/tp-retrieve &$\dcloudID$/tp-retrieve\\
\cline{2-3}
&$\DockerHubID$/tp-decompress &$\dcloudID$/tp-decompress\\
\cline{2-3}
&$\DockerHubID$/tp-ha-train &$\dcloudID$/tp-ha-train\\
\cline{2-3}
&$\DockerHubID$/tp-la-train &$\dcloudID$/tp-la-train\\
\cline{2-3}
&$\DockerHubID$/tp-ha-score &$\dcloudID$/tp-ha-score\\
\cline{2-3}
&$\DockerHubID$/tp-la-score &$\dcloudID$/tp-la-score\\
\hline
\end{tabular}}
\end{table}
\subsection{Docker registry services}
This work uses two different Docker registries to store, manage, and share the Docker images: \textit{1)} the public Docker Hub\textsuperscript{\ref{footnot:dock-hub}}, and \textit{2)} a MinIO-based Docker registry locally deployed in our laboratory\footnote{\url{https://dcloud2.itec.aau.at:9001}\label{footnot:aau-hub}}. MinIO as an open-source and distributed object storage platform\footnote{\url{https://github.com/minio/minio}} supports cloud object storage technologies such as Amazon S3
, enabling the storage of unstructured data such as multimedia, log files, and container or virtual machine images. The regional MinIO-based Docker registry is provisioned on a local server with a specific storage capacity according to the user's requirements (e.g., \qty{100}{\giga\byte})\footnote{\url{https://github.com/SiNa88/docker-registry}}.
\paragraph*{Docker images} 
Table~\ref{tab:docker-img} summarizes the image IDs available at two Docker registries: the public Docker Hub and a regional one at the Edge\textsuperscript{\ref{footnot:aau-hub}}. We used the official images of \texttt{amd64/ubuntu:18.04}, \texttt{ubuntu:24.10
}, \texttt{alpine:3}, \texttt{python:3.9-slim}, and \texttt{python:3.9} to create our Docker images for different microservices. 
Afterward, we tagged the images with the labels \texttt{amd64} (for hardware architectures \texttt{x86} and \texttt{amd}) and \texttt{arm64} (for \texttt{arm} architectures).

\subsection{Benchmarks of microservice executions} 
\begin{table}[!t]
\centering
\caption{Benchmarks of 
microservices deployed from Docker Hub and regional
registries, and executed on 
edge devices.}
\label{tbl:benchmark:dep+exec+docker+hub}
\resizebox{\columnwidth}{!}
{ 
\begin{tabular}{|c||cccccc|}
\cline{2-7}
\multicolumn{1}{c}{} &\multicolumn{1}{|c|}{\makecell{\textit{Micro-}\\\textit{service} ($m_i$)}}  & \multicolumn{1}{c|}{\makecell{$\SIZE_{m_i}$\\$[\si{\giga\byte}]$}}   & \multicolumn{1}{c|}{$T_p\ [\si{\second}]$} & \multicolumn{1}{c|}{$\ct\ [\si{\second}]$} &\multicolumn{1}{c|}{\makecell{$\ec\ [\si{\joule}]$\\\texttt{medium}}}  & \multicolumn{1}{c|}{\makecell{$\ec\ [\si{\joule}]$\\\texttt{small}}} 
\\ \hline
\multirow{6}{.4cm}{\centering {\rotatebox[origin=c]{90}{\makecell{{{\textit{Video}}}\\{{\textit{processing}}}}}}}&\multicolumn{1}{c|}{\textit{Transcode}}&\multicolumn{1}{c|}{0.17}  & \multicolumn{1}{c|}{17.5--19}   &\multicolumn{1}{c|}{82--85}  & \multicolumn{1}{c|}{856--859} & \multicolumn{1}{c|}{340--355}
 \\ \cline{2-7}
&\multicolumn{1}{c|}{\textit{Frame}}              & \multicolumn{1}{c|}{0.70} & \multicolumn{1}{c|}{10--20} & \multicolumn{1}{c|}{147--184} & \multicolumn{1}{c|}{355--378} & \multicolumn{1}{c|}{557--679} 
\\ \cline{2-7}
&\multicolumn{1}{c|}{\textit{HA Train}} &\multicolumn{1}{c|}{5.78} & \multicolumn{1}{c|}{121--124} & \multicolumn{1}{c|}{1071--1421} & \multicolumn{1}{c|}{3240--3288} & \multicolumn{1}{c|}{4654--5472}  
\\ \cline{2-7}
&\multicolumn{1}{c|}{\textit{LA Train}}  &
\multicolumn{1}{c|}{5.78} & \multicolumn{1}{c|}{87--97} & \multicolumn{1}{c|}{1058--1297} & \multicolumn{1}{c|}{1834--1849}&  \multicolumn{1}{c|}{3995--4700}   
\\ \cline{2-7}
&\multicolumn{1}{c|}{\textit{HA Infer}} &\multicolumn{1}{c|}{3.53} & \multicolumn{1}{c|}{38--41} & \multicolumn{1}{c|}{356--435} & \multicolumn{1}{c|}{849--850} & \multicolumn{1}{c|}{1423--1602} \\ \cline{2-7}
&\multicolumn{1}{c|}{\textit{LA Infer}} & \multicolumn{1}{c|}{3.54} & \multicolumn{1}{c|}{38-40}  & \multicolumn{1}{c|}{350--429} & \multicolumn{1}{c|}{819--842} & \multicolumn{1}{c|}{1400--1590} \\
\hline
\multirow{6}{.4cm}{\centering {\rotatebox[origin=c]{90}{\makecell{{{\textit{Text}}} \\{{\textit{processing}}}}}}}&\multicolumn{1}{c|}{\textit{Retrieve}} &\multicolumn{1}{c|}{0.14}  & \multicolumn{1}{c|}{42--58}  & \multicolumn{1}{c|}{331--334}  & \multicolumn{1}{c|}{144--173}  & \multicolumn{1}{c|}{1136--1183}  
\\ \cline{2-7}
&\multicolumn{1}{c|}{\textit{Decompress}} &\multicolumn{1}{c|}{0.78}&  \multicolumn{1}{c|}{27--55} & \multicolumn{1}{c|}{290--331} & \multicolumn{1}{c|}{415--432} & \multicolumn{1}{c|}{1037--1143}\\ \cline{2-7}
&\multicolumn{1}{c|}{\textit{HA Train}} &\multicolumn{1}{c|}{2.36} & \multicolumn{1}{c|}{139--144} &\multicolumn{1}{c|}{427--507} & \multicolumn{1}{c|}{3482--3728} & \multicolumn{1}{c|}{1638--1903} \\ \cline{2-7}
&\multicolumn{1}{c|}{\textit{LA Train}} &  \multicolumn{1}{c|}{2.36} & \multicolumn{1}{c|}{87--89} & \multicolumn{1}{c|}{288--363}  & \multicolumn{1}{c|}{1622--1642}& \multicolumn{1}{c|}{870--985} \\ \cline{2-7}
&\multicolumn{1}{c|}{\textit{HA Score}} & \multicolumn{1}{c|}{0.63} & \multicolumn{1}{c|}{74--76} & \multicolumn{1}{c|}{177--211} & \multicolumn{1}{c|}{1228--1319}& \multicolumn{1}{c|}{675--786} \\\cline{2-7}
&\multicolumn{1}{c|}{\textit{LA Score}} &\multicolumn{1}{c|}{0.63} & \multicolumn{1}{c|}{75--78} &  \multicolumn{1}{c|}{175--210} & \multicolumn{1}{c|}{1295--1299}    & \multicolumn{1}{c|}{670--785}\\ 
\hline
\end{tabular}}
\end{table}
Table~\ref{tbl:benchmark:dep+exec+docker+hub} summarizes the benchmarks on the devices for all microservices involved in the two case study applications. We measured the energy consumption of the \texttt{Intel}-based devices using \texttt{pyRAPL}\footnote{\url{https://github.com/powerapi-ng/pyRAPL}} and the \texttt{ARM}-based ones using a 
Ketotek power meter. We also used the \texttt{date} command\footnote{\url{https://man7.org/linux/man-pages/man1/date.1.html}} within a \texttt{shell script} to measure each microservice deployment time alongside dataflow processing and transmission times.

\section{Experimental Results} \label{sec:experimntresult}
\begin{table}[!t]
\centering
\caption{Distribution (in \si{\percent}) of Docker image deployments from registries and executions on devices for two case studies.}
\label{tab:distributions}
\resizebox{.8\columnwidth}{!}
{
\begin{tabular}{|c||c||c|c|}
\cline{2-4}\multicolumn{1}{c|}{}& \emph{Device} & \makecell{\emph{Docker} \emph{Hub}} & \makecell{\emph{Regional} \emph{Registry}}\\
\hline
\multirow{2}{*}{\centering {{\makecell{{{\textit{Video}}} {{\textit{processing}}}}}}}
& \texttt{medium}  &  \num{83} \%  &  --     \\\cline{2-4}
& \texttt{small}   &  --  &  \num{17} \%    \\
\hline\hline
\multirow{2}{*}{\centering {{\makecell{{{\textit{Text}}} {{\textit{processing}}}}}}}
& \texttt{medium}  &  \num{17} \%  &  \num{17} \%     \\\cline{2-4}
& \texttt{small}   &  --  &  \num{66} \%      \\
\hline
\end{tabular}
}
\end{table}
Table~\ref{tab:distributions} shows the distributions (in \si{\percent}) of the microservices retrieved from Docker registries and executed on edge computing devices. 
\begin{figure}[!t]
    \centering
    \subfloat[\centering Energy consumed by each microservice executed on an Edge device scheduled by \deep{}.]{\includegraphics[width=.49\columnwidth]{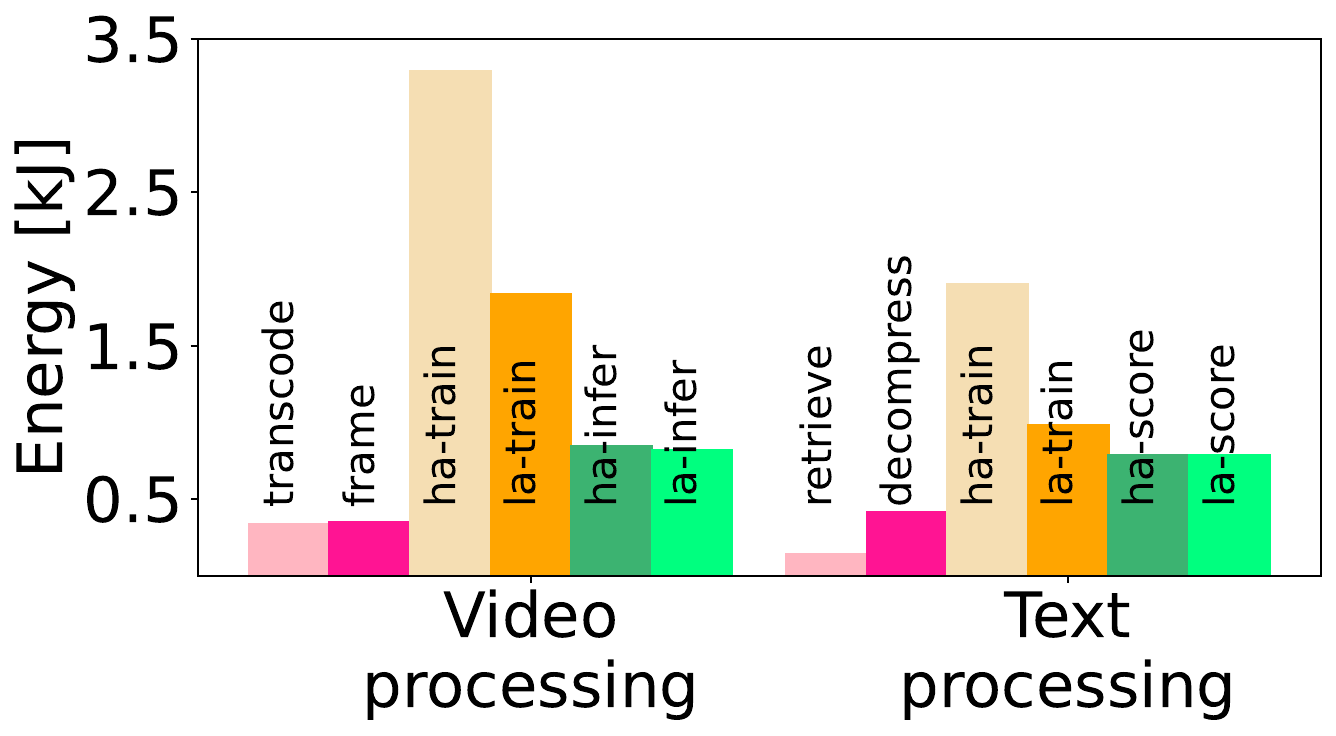}\label{fig:eval_energy1}}
    \hfill
    \subfloat[\centering Energy consumed using three different microservice image deployment methods.]{\includegraphics[width=.49\columnwidth]{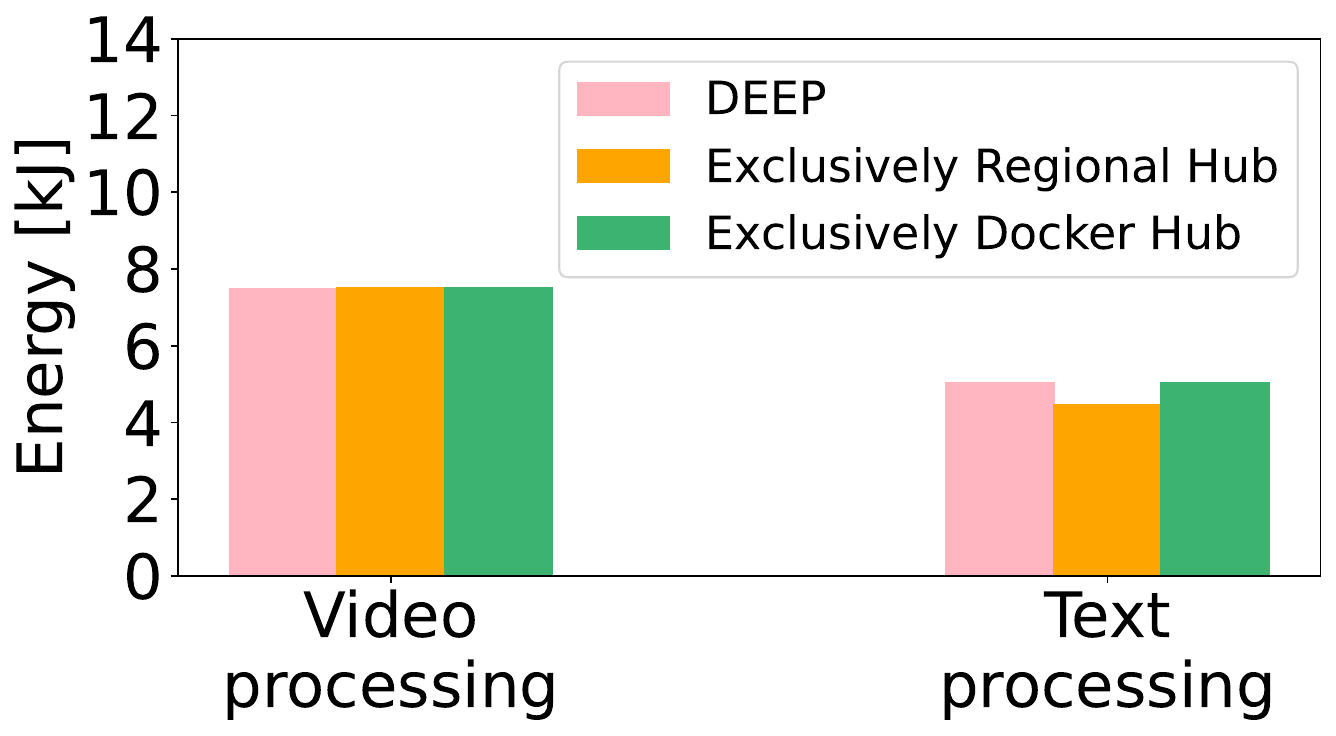}\label{fig:eval_energy2}}
    \caption{Energy consumption (two case studies).}
\end{figure}
Figure~\ref{fig:eval_energy1} shows that HA and LA training microservices of both applications consume more energy compared to other ones. Moreover, Figure~\ref{fig:eval_energy2} compares the energy performance by dataflow processing \deep{} with exclusively Docker Hub and regional registry methods. The results show that \deep{} reduces the energy consumption of video processing by \SI{.2}{\percent} ($\approx$\SI{14}{\joule}) compared to two other methods. Moreover, it reduces the text processing energy consumption by \SI{.34}{\percent} ($\approx$\SI{18}{\joule}) compared to the exclusively Docker hub method. In other words, the regional Docker registry shows competitive energy efficiency compared to Docker Hub.

\section{Conclusion} \label{sec:conclusion}
We presented \deep{}, an edge-based dataflow processing method deploying microservices from hybrid Docker Hub and regional registries. \deep{} enables the reduction of edge devices' energy consumption and yields more energy-efficient dataflow processing, especially in the text processing application. The results show that \deep{} reduces energy consumption by \SI{0.34}{\percent}, compared to exclusively deploying from Docker Hub, by retrieving \SI{83}{\percent} of the text processing microservices from the regional registry. 
We also observe that the microservice's image location plays no significant role in energy consumption for heavyweight video processing.
We plan to extend this energy-aware nash-based model to schedule the computation between cloud and edge.

\section*{Acknowledgment} This work received funding from EXDIGIT 
project funded by Land Salzburg, grant number 20204-WISS/263/6-6022, and \mbox{Graph-Massivizer} by \emph{European Union}'s Horizon research and innovation 
program, grant agreements \num{101093202}.

\bibliographystyle{unsrt}
\bibliography{references}

\end{document}